\newcommand{\eqn}[1]{(\ref{#1})}

\documentclass[aps,prb,twocolumn,superscriptaddress,showpacs]{revtex4}

\usepackage{graphicx}

\begin{document}
\newcommand{\ket}[1]{\left|#1\right\rangle}
\newcommand{\bra}[1]{\left\langle#1\right|}
\newcommand{\im}{i}
\newcommand{\imag}{\grave{\imath}}

\title{A Chain-Boson Model for the Decoherence and Relaxation \newline of a Few Coupled SQUIDs in a Phonon Bath}

\author{A. J. Skinner}
\affiliation{Department of Physics, Skidmore College, Saratoga Springs, NY 12866}
\affiliation{Joint Quantum Institute, Department of Physics, University of Maryland, College Park, MD 20740}
\author{B.-L. Hu}
\affiliation{Joint Quantum Institute, Department of Physics, University of Maryland, College Park, MD 20740}

\date{\today}

\begin{abstract}
We develop a ``chain-boson model'' master equation, within the Born-Markov approximation, for a few superconducting quantum interference devices (SQUIDs) coupled into a chain and exchanging their angular momenta with a low temperature phonon bath.  Our master equation has four generators; we concentrate on the damping and diffusion and use them to study the relaxation and decoherence of a Heisenberg SQUID chain whose spectrum exhibits critical point energy-level crossings, entangled states, and pairs of resonant transitions.  We note that at an energy-level crossing the relevant bath wavelengths are so long that even well-spaced large SQUIDs can partially exhibit collective coupling to the bath, dramatically reducing certain relaxation and decoherence rates.  Also, transitions into entangled states can occur even in the case of an independent coupling of each SQUID to the bath.  Finally, the pairs of  resonant transitions can cause decaying oscillations to emerge in a lower energy subspace.
\end{abstract}


\maketitle

\section{Introduction}
\subsection{SQUIDs and SQUID Chains}
A superconducting quantum interference device (SQUID) can be made from a small strip of aluminum bent into a ring, joined at the ends, and cooled to a \mbox{milli}Kelvin temperature.  Aluminum is a superconductor and an aluminum oxide layer, where the ends meet, forms a ``Josephson Junction'' potential barrier.  Precise tuning of an externally imposed magnetic flux can cause oscillations of current \cite{Friedman, vanderWal} between clockwise ``$\ket{\uparrow}$'' and counterclockwise ``$\ket{\downarrow}$'' states via an evolving phase interference, $e^{\im(E_1-E_0)t/\hbar}$, between the ground state, $\ket{0} \equiv (\ket{\uparrow}+\ket{\downarrow})/\sqrt{2}$, and first excited state, $\ket{1} \equiv(\ket{\uparrow}-\ket{\downarrow})/\sqrt{2}$, which span the low-energy dynamics of the device and form a logical basis for a ``qubit'' of quantum information, $c_0 \ket{0}+c_1\ket{1}$.

Many SQUIDs can be coupled together into a chain.  The aluminum rings are not topologically linked but their proximity allows capacitive and inductive interaction between nearest neighbor SQUIDs \cite{Levitov,Lyakhov}.  We are principally interested in using the chain to encode  \cite{LidarWu,DiVincenzo,BaconBrownWhaley} and protect or correct \cite{Kribs,Duan,Lidar,Zanardi} quantum information.  ``Logical gate'' operations on a chain of SQUIDs can create \cite{Berman} useful non-classical correlations quantified by their ``entanglement of formation'' \cite{Wootters}.  Chains can propagate excitations, qubits \cite{OsborneLinden, Subrahmanyam,ChristandlLandahl,BurgarthBose}, and even entangled singlets, $(\ket{\uparrow\downarrow}-\ket{\downarrow\uparrow})/\sqrt{2}$ \cite{Cubitt,PrattEberly,Amico}.  They could also provide long-sought experimental realizations of spin chains which in turn illustrate the correlations and phase transitions of many-body physics.  Cooling a chain to its zero-temperature ground state can prepare useful entanglement \cite{BrussDatta, Huelga}.  Ground state entanglement can vanish abruptly, for example, as spin-spin couplings are adjusted across a critical point \cite{Osterloh,Osborne,WuLidar}.  In that case, it is the intermediate energy states that can possess entanglement; their quantum correlations can be evident, at a warmer but not too-high temperature, when mixed sufficiently into the equilibrium state \cite{Arnesen,Asoudeh}.

\begin{figure}
\begin{center}
\includegraphics{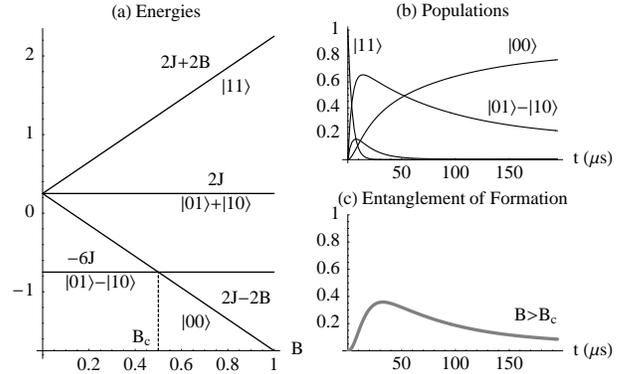}
\end{center}
\caption{Relaxing through entangled states.  A Heisenberg two-SQUID chain above the critical point $B_c$ cooling from $\ket{11}$ to $\ket{00}$.  In the process, the states $\ket{01}\pm\ket{10}$ are occupied, resulting in a surge of the entanglement of formation, even though the ground state is separable and the large SQUIDs are dissipating independently.   ($8J/h=1.0$GHz, $2B/h=1.5$GHz, $k_BT/h=0.3$GHz, $R=10 \mu$m, $I=3 \mu$A, $\rho=5$g$/$cm${}^3$, $c_{\perp}=5$km$/$s).}
\label{2SQUIDsFreqPopsEof}
\end{figure}

\subsection{Decoherence and Relaxation}
Here we consider some effects of exposing a SQUID chain to a low temperature bath of phonons.  We suppose the SQUIDs are lithographically etched and deposited into a solid crystal (e.g.\ of silicon) which, for simplicity, we assume surrounds the chain.  In a SQUID's oscillation between current states, the conservation of total angular momentum requires torsional oscillations of the solid and thus the emission and absorption of phonons \cite{Chudnovsky2}.  Decoherence of superconducting qubits is also caused by two-level defects in the Josephson Junction barrier material \cite{Martinis}, which could in principle be purified, whereas here we consider only the minimal unavoidable coupling required by symmetry.

Usually, when a quantum system is opened-up to its finite-temperature environment, its energies and eigenstates are perturbed and the new energies and stationary states are viewed as renormalized quantities.  Then it equilibrates, by decoherence and relaxation, to a stationary thermal mixture of these eigenstates.  In the relaxation process, the populations (i.e.\ probabilities) of each eigenstate are adjusted until a thermal mixture is obtained.  Decoherence is the decay of the oscillations (e.g.\ of the current).  Some decoherence goes along with the relaxation --- adjusting the probabilities of the eigenstates undermines the support for any phase interference between them --- but it can also be caused by pure dephasing, in which the phase's probability distribution spreads, reducing the average phase interference, or ``coherence,'' without adjusting the populations.  As described in appendix~\ref{cohpop}, the populations and coherences are, respectively, the diagonal and off-diagonal matrix elements, in the energy eigenbasis, of the chain's density operator $\rho(t)$.  The absolute value of a coherence is just the (decaying) envelope of the oscillations.

We are interested in large SQUIDs of radius $R=10 \mu$m with a $3
\mu$A current oscillating at $1.0 \pm 0.5$ GHz.  For a single SQUID
the decoherence model gives decoherence times of a few $\mu$s
\cite{Chudnovsky2} which are not inconsistent with those of recent
experiments \cite{Friedman, Mooij}.  On the other hand we will also
simulate small SQUIDs, with $R=10$ nm and $I=0.1 \mu$A, to
demonstrate the concept of a collective coupling to the environment.
The concept of a collective coupling is relevant to the critical
point open system dynamics of the large SQUIDs.

\subsection{The Chain-Boson Model}
The spin-boson model \cite{Leggett} has been widely applied towards a better understanding of the environment's effect on a single qubit's coherence.  In the spin-boson model, a two-level system is coupled to an environment of oscillators which model a heat bath, such as is often used in studies of quantum Brownian motion \cite{Ford,HPZ92,Primer}.  Because of the coupling, the system becomes entangled with the bath.  When averaged over environmental outcomes, the system typically loses coherence and thermalizes.  Chains of harmonic  oscillators interacting with a finite temperature bath have been considered in \cite{HMM03,Eisert,CYH}.

\begin{figure}
\begin{center}
\includegraphics{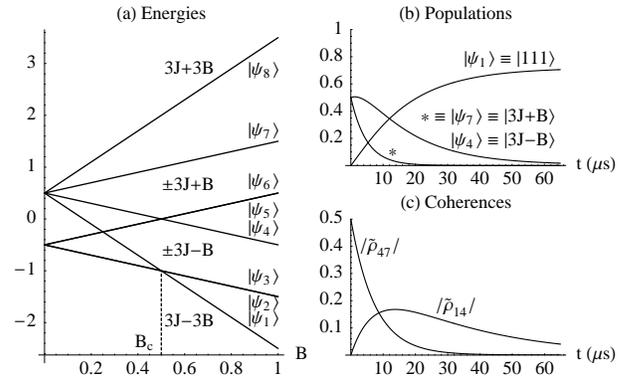}
\end{center}
\caption{Decaying oscillations emerging in a lower energy subspace.  Three large SQUIDs dissipating independently.  $\ket{\psi_4}+\ket{\psi_7}$ cooling through a mixture including $\ket{\psi_1}+\ket{\psi_4}$.  The transitions $\ket{\psi_4}\leftrightarrow\ket{\psi_1}$ and $\ket{\psi_7}\leftrightarrow\ket{\psi_4}$ are resonant.  $\tilde{\rho}_{47}$, which decays $\sim e^{-\bar{\Gamma}_{47} t}$, is absorbed into $\tilde{\rho}_{14}$ which decays more slowly $\sim e^{-\bar{\Gamma}_{14}t}$.  It might be easier to start with $\ket{\uparrow\uparrow\uparrow}$, close to the equal superposition of $\ket{\psi_4} = \ket{001}+\ket{100}+\ket{010}$ and $\ket{\psi_7} = \ket{011} +\ket{110} +\ket{100}$, to prepare the coherence $\tilde{\rho}_{47}$ which flows into $\tilde{\rho}_{14}$.  ($6J/h = 1.0$GHz, $2B/h=1.5$GHz, $k_BT/h=0.3$GHz, $R=10 \mu$m, $I=3 \mu$A).}
\label{3SQUIDsCohFlow}
\end{figure}

In light of qubit encoding schemes and the desire to process and protect quantum information, it is necessary to study the decoherence of multiple qubits.  In the chain-boson model one embeds a chain of qubits in a bosonic bath so that the qubits experience a {\it{location-dependent}} interaction with the bath variables.

For a system comprising a register of qubits, two types of system-bath coupling have already been extensively considered.  The simplest is a collective coupling, in which each qubit couples to the same environmental variable as the rest.  This is appropriate to scenarios where the qubits are spaced closer together than the relevant wavelengths of the bath, i.e.\ those corresponding to the qubits' transition frequencies.  The symmetries of the collective coupling can lead, with certain system Hamiltonians, to decoherence free subspaces \cite{Lidar,DuanGuo,Zanardi} and Dicke super- and sub-radiance and super- and sub-decoherence \cite{DickeSuper,ChudnovskySuper}.

The other commonly used type of system-bath coupling is the independent coupling model, in which each qubit couples to its own bath, separate from the baths used for the other qubits.  This is an appropriate model for qubits spaced farther apart than the relevant wavelengths of the bath.  And in the context of solid state qubits, a significant source of noise is the voltage leads that control the qubits.  With one lead per qubit, the independent baths model is a natural assumption.  The independent coupling model has been used to study the decoherence during two-qubit logic gates \cite{Thorwart,Ahn} as well as the entanglement rate for coupled qubits \cite{YiWang,Huelga,WangBatelaan}.

When there is one voltage lead controlling multiple qubits, one typically uses the collective coupling model, as each qubit is experiencing the same electronic noise.  A likely scenario for a pair of qubits is an independent lead for each qubit as well as one common lead.  The disentanglement and decohering of the two qubits in two separate and one common cavity have been considered \cite{YuEberly1,YuEberly2,YuEberly3,FicTan1,FicTan2,ASH}.  Several works compare the collective and independent bath scenarios for coupled qubits \cite{Governale,Storcz2003,Abliz,Storcz2005}.  Fine-tuning the inter-qubit coupling to protect against collective dissipation has also been studied \cite{Storcz2005A, Storcz2005B}.  In comparison, the optimum qubit-qubit coupling was examined for the case of independent dissipation \cite{Grigorenko,YouHuNori}.

Most studies fall into these two categories, whether the qubits are coupled together or not.  But for uncoupled qubits there has been careful consideration of the intermediate scenario, in which the qubits are neither far apart nor close together \cite{PalmaEkert,Johnson,Ischi}.  The relaxation and decoherence rates depend on $\vec{k}\cdot \vec{r}_{jk}=\omega \, \tau_{jk}$, where the $\vec{k}$ are the bath wavevectors that interact with the register of qubits at its transition frequency $\omega$, and $\tau_{jk}$ is the phonon transit time between qubits $j$ and $k$ separated by $\vec{r}_{jk}$.  These results make an elegant transition between the two limiting cases of qubits close together and far apart.  In the intermediate scenario the bath can induce entanglement between uncoupled qubits \cite{Piani}, as can also happen with a collective coupling to the bath \cite{Braun}.  Uncoupled SQUIDs can also be entangled when the bath is specialized to a single mode cavity \cite{Zhang}.

\subsection{Role of the Critical-Point}
In our chain of SQUIDs coupling to phonons, an essential point is that the inter-qubit couplings can play a crucial role in determining which are the relevant bath wavevectors and thus whether the qubits couple collectively or independently to the bath.  In our model, there is a critical point energy-level crossing at which some transition frequencies are so slow that $\vec{k}\cdot \vec{r}_{jk}= \omega \, \tau_{jk}\ll1$ even though the qubits may be far apart with respect to the uncoupled qubits' transition frequencies.  At the critical point the chain can thus obtain some or all of the benefits of a collective coupling to the bath.  Some interesting studies include the intermediate regime for coupled qubits \cite{Dube,Pasquale} and detail the sub-radiant behavior of dipole-coupled qubits for small $\vec{k}\cdot\vec{r}_{ij}$ \cite{Ficek}, but the role of the inter-qubit coupling in determining the relevant bath wavevectors and their relation to the inter-qubit distance is not emphasized.

\subsection{Overview}
Working from the total Hamiltonian $H=H_S+H_B+V$, describing the
SQUIDS, the bath, and their coupling, we apply master equation
techniques from the quantum Brownian motion model
\cite{Ford,HPZ92,Primer}, in the Born-Markov approximation, to a
chain of SQUIDs interacting with a phonon heat bath via
location-dependent couplings.  The resulting generators of the open
system dynamics are associated with four types of coefficients: the
renormalization, anomalous diffusion, damping, and diffusion.  We use
the damping and diffusion to develop a matrix element equation for
the populations and coherences.  It is similar to a Bloch-Redfield
\cite{Redfield} equation but in the interaction picture.  It gives
the relaxation and decoherence rates, as well as the possibility for
coherent oscillations to move \cite{BarnesWarren} from one subspace,
where they are decaying, into a lower energy subspace with a longer
decoherence time.  A complete network of selection-ruled transitions
leads to thermalization.  This is usually the case for large
well-spaced SQUIDs.  For small closely-spaced SQUIDs the network is
broken by the degeneracies of their collective coupling to the bath;
decoherence and transition rates can scale with the number of SQUIDs
or vanish; some subspaces are protected.  Finally, at the critical
point even the large well-spaced SQUIDs can acquire some of this
collective Dicke super- and sub-radiant and super- and sub-decoherent
behavior.

\begin{figure}
\begin{center}
\includegraphics{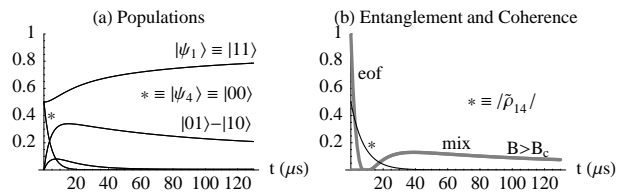}
\end{center}
\caption{Decoherence and Relaxation in two large SQUIDs dissipating independently.  $\ket{00}+\ket{11}$ cooling above the critical point.  (a) The population dynamics are the same for an equal mixture of $\ket{\psi_1}\equiv\ket{11}$ and $\ket{\psi_4}\equiv\ket{00}$; only the superposition is entangled.  (b) The entanglement does not oscillate because any phase between $\ket{00}$ and $\ket{11}$ may be generated locally.  It decays faster than the coherence  $\tilde{\rho}_{14} \sim e^{-\bar{\Gamma}_{14}t}$ upon which it depends.  Later, the populations mix back in some entanglement. ($8J/h = 1.0$GHz, $2B/h=1.5$GHz, $k_BT/h=0.3$GHz, $R=10 \mu$m, $I=3 \mu$A).}
\label{2SQUIDsDecohRelax}
\end{figure}

\section{The Chain, Bath, and their Coupling}
\subsection{The Heisenberg SQUID Chain}
For its mathematical simplicity and relevance to quantum information processing, we consider the isotropic Heisenberg coupling between nearest neighbor SQUIDs.  In principle this can be engineered with a precise balance of inductive and capacitive coupling between nearest-neighbor SQUIDs.  However, several of our methods are applicable to other types of coupling.  The ``antiferromagnetic'' ($J>0$) Heisenberg chain of $N$ SQUIDs evolves by its Hamiltonian,
\begin{equation}
H_S = \sum_{j=1}^{N} (J \vec{\sigma}_j\cdot\vec{\sigma}_{j+1} - B \sigma_j^x),
\label{H_S}
\end{equation}
assuming periodic boundary conditions $\vec{\sigma}_{N+1}\equiv\vec{\sigma}_1$.  Here, $\vec{\sigma}_j$ are the Pauli matrices for the $j$th SQUID, with $\ket{\uparrow}$ and $\ket{\downarrow}$ the eigenstates of $\sigma^z$, and $B$ is an energy splitting (not an imposed magnetic field) appropriate to the natural precession of our pseudo-spin qubits: coherent quantum oscillations between $\ket{\uparrow}$ and $\ket{\downarrow}$.

The Heisenberg and magnetic sums commute and their respective quantum numbers $l$ and $m$ determine the energy spectrum $\{l J - m B\}$ (up to some degeneracies not split by J and B).  Regardless of $N$, each eigenstate is a linear combination of states with the same number of $\ket{1}$ vs.\ $\ket{0}$ qubits ($m\equiv N_0-N_1$) and is typically entangled, with the exception of the extremal $m$ states $\ket{11\ldots}$ and $\ket{00\ldots}$.  Increasing $B$ relative to fixed $J$ causes energy-level crossings.  At a critical point, $B_c$, the ground state changes from entangled to unentangled.

The two-SQUID chain's energies are shown in Figure~\ref{2SQUIDsFreqPopsEof} as a function of the single-SQUID energy splitting $B$.  Figure~\ref{3SQUIDsCohFlow} plots the energies for a three-SQUID chain.  Their respective eigenstates are as follows:
\[
\begin{array}{lcr}
\begin{array}{l}
{\mbox{Two-SQUID Chain}} \\
\hline
\ket{\psi_1} \equiv \ket{00} \\
\ket{\psi_2} \equiv \ket{01}-\ket{10} \\
\ket{\psi_3} \equiv \ket{01}+\ket{10} \\
\ket{\psi_4} \equiv \ket{11} \\
\end{array}
& &
\begin{array}{l}
{\mbox{Three-SQUID Chain}} \\
\hline
\ket{\psi_1} \equiv \ket{000} \\
\ket{\psi_2} \equiv \ket{001}-\ket{100} \\
\ket{\psi_3} \equiv \ket{001}+\ket{100}-2\ket{010} \\
\ket{\psi_4} \equiv \ket{001}+\ket{100}+\ket{010} \\
\ket{\psi_5} \equiv \ket{011}-\ket{110}\\
\ket{\psi_6} \equiv \ket{011}+\ket{110}-2\ket{101} \\
\ket{\psi_7} \equiv \ket{011} +\ket{110} +\ket{101} \\
\ket{\psi_8} \equiv \ket{111}.\\
\end{array}
\end{array}
\]

\subsection{The Harmonic Phonon Bath}
The harmonic crystal Hamiltonian is composed of phonon modes labelled by wavevector $\vec{k}$ and polarization index $s$.  Each phonon contributes an energy $\hbar \omega_s(\vec{k})$:
\begin{equation}
H_B=\sum_{\vec{k},s}\hbar \omega_s(\vec{k}) a^{\dagger}_{\vec{k}s}a_{\vec{k}s}.
\label{H_B}
\end{equation}
Here $a^{\dagger}_{\vec{k}s}$ and $a_{\vec{k}s}$ are the phonon creation and annihilation operators.  For frequencies below a ``Debye'' or cutoff frequency $\Lambda$ we assume linear dispersions, $\omega_{1,2}(\vec{k}) = c_{\perp}\,  |\vec{k}|$ and $\omega_3(\vec{k}) = c_{\parallel}\,  |\vec{k}|$, for transverse ($\perp$) and longitudinal ($\parallel$) polarizations $\hat{e}_s(\vec{k}) = \left\{\hat{k}_{\perp 1}, \hat{k}_{\perp 2}, \hat{k}_{\parallel}\right\}$.  We will also make use of the momentum density at site $\vec{r}$:
\begin{equation}
\vec{\pi}(\vec{r}) =-\im \sum_{\vec{k}s}\sqrt{\hbar \omega_s(\vec{k}) \rho} \; \frac{a_{\vec{k}s}e^{\im \vec{k}\cdot\vec{r}}-a^{\dagger}_{\vec{k}s}e^{-\im \vec{k}\cdot\vec{r}}}{\sqrt{2 V}} \; \hat{e}_s(\vec{k})
\label{momentumdensity}
\end{equation} with $V$ and $\rho$ the volume and mass density of the crystal.

\begin{figure}
\begin{center}
\includegraphics{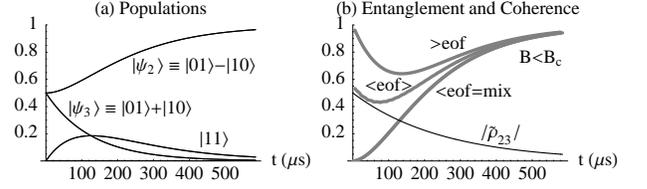}
\end{center}
\caption{Entanglement oscillations in two large SQUIDs dissipating independently.  $\ket{01}$ cooling below the critical point.  (a) The population dynamics are the same for an equal mixture of $\ket{\psi_2} \equiv \ket{01} - \ket{10}$ and $\ket{\psi_3} \equiv \ket{01} + \ket{10}$.  (b) From their initial superposition $\ket{01}$, the evolving phase between $\ket{\psi_2}$ and $\ket{\psi_3}$ drives $\approx 6 \times 10^5$ rapid oscillations of the entanglement of formation (eof); we have plotted a moving average $<$eof$>$, the upper bound ($>$eof), and the lower bound ($<$eof) which is the same as the entanglement of the mixture.  Initially, the moving average decays with the coherence $\tilde{\rho}_{23} \sim e^{-\bar{\Gamma}_{23}t}$ but later pulls away to equilibrate.  ($8J/h = 1.0$GHz, $2B/h=0.5$GHz, $k_BT/h=0.1$GHz, $R=10 \mu$m, $I=3 \mu$A).}
\label{2SQUIDsEofOsc}
\end{figure}

\subsection{The Chain-Bath Coupling}
The coupling between the SQUIDs and the crystal arises from the fact that each SQUID's current $I \sigma_j^z$ is formed from the electronic band states in the reference frame co-moving with the lattice sites during the torsional oscillations of the crystal \cite{Chudnovsky2}.  In the lab frame the electron velocity $\vec{v}_e$ must include the speed $\dot{\vec{u}}=\vec{\pi}/\rho$ of the lattice sites: $\vec{v}_e = \vec{j}/(e n_e)+\dot{\vec{u}}$.  Here $\vec{j}$ is the current density and $e$ and $n_e$ are the electron charge and number density of the electrons.  Their kinetic energy density $n_e \frac{1}{2}  m_e |\vec{v}_e|^2$ thus acquires a cross term and the total Hamiltonian must include an additional
\begin{equation}
V = (m_e/e) \int d^3r  \, \vec{j}\cdot\dot{\vec{u}}\,.
\label{kineticenergydensity}
\end{equation}

In appendix~\ref{chainbathcoupling} we first derive, following \cite{Chudnovsky2}, the coupling $V_j$ of an individual SQUID to the crystal and then, because the current density $\vec{j}$ is the sum of the individual densities $\vec{j}_j$, sum their contributions at SQUID locations $\vec{x}_j$ (spaced a distance $d$ apart along the $x$-axis) to obtain
\begin{equation}
V = \lambda \sum_j {\cal{X}}_j\, \phi_j
\label{bilinearV}
\end{equation}
akin to a quantum Brownian oscillator's bilinear coupling to an oscillator bath, $\lambda x \phi$, (only here we are summing over several contact-points between the chain and the bath).  The coupling constant $\lambda \equiv  2 \pi I (m_e/e) \sqrt{R \hbar c_{\perp}/2 \rho V}$ fixes the strength of the interaction between chain operators ${\cal{X}}_j \equiv \sigma_j^z$ and bath operators
\begin{equation}
\phi_j \equiv \sum_{\vec{k}} \sqrt{|\vec{k}| R}  \, J_1(k_x R) \left(a_{\vec{k}1}e^{\im \vec{k}\cdot \vec{x}_j}  + a^{\dagger}_{\vec{k}1}e^{-\im \vec{k}\cdot \vec{x}_j}\right),
\end{equation}
for which we have defined the first transverse polarization $\hat{e}_1(\vec{k})$ to lie in the plane of the SQUID ring and $J_1(k_x R) = J_1(|\vec{k}|R  \sin \theta)$ is the first order Bessel function (and $\theta$ is the polar angle between the $\hat{z}$-axis of the ring and the wavevector $\vec{k}$).

The selection rules, $\bra{l'm'}{\cal{X}}_j\ket{lm} \sim \delta_{m',m\pm2}$ reflect the fact that the ``interaction operator'' ${\cal{X}}_j$ simply flips the $j^{\mbox\scriptsize th}$ $\ket{0}$ or $\ket{1}$, giving a non-zero probability only to obtain an eigenstate with one more $\ket{1}$ or $\ket{0}$.  The allowed bath-driven transition frequencies are $\omega = [ (l-l')J \pm 2 B]/\hbar$, and transitions driven between degenerate states are thus between distinct $l$ and $l'$ at a crossing of energy-levels.

The two-SQUID chain's interaction operators are, in its energy eigenbasis,
\begin{equation}
{\cal{X}}_{1,2}=\frac{1}{\sqrt{2}}\left(\begin{array}{cccc}
0 & \mp1 & 1 & 0 \\
\mp1 & 0 & 0 & \pm1 \\
1 & 0 & 0 & 1 \\
0 & \pm1 & 1 & 0 \\
\end{array}\right)
.
\end{equation}
They present a ``network'' of selection-ruled transitions: $\ket{\psi_1} \leftrightarrow \ket{\psi_2} \leftrightarrow \ket{\psi_4}$ and $\ket{\psi_1} \leftrightarrow \ket{\psi_3} \leftrightarrow \ket{\psi_4}$ with transition energies $2B$ and $2B\pm8J$.

For future reference, this network should be compared with the case of a collective interaction operator  $J_z= \sum_j {\cal{X}}_j$ in which (for two SQUIDs) the $\pm1$ matrix elements above cancel and the selection-rules constrict to $\ket{\psi_1} \leftrightarrow \ket{\psi_3} \leftrightarrow \ket{\psi_4}$ leaving $\ket{\psi_2}$ as a protected ($1$-dimensional) subspace.

\section{Formalism}
\subsection{The Born Approximation}
We now apply master equation techniques from the model of quantum Brownian motion \cite{Ford,Primer, HPZ92} to the chain of SQUIDs interacting with their phonon bath.

The unperturbed Hamiltonian $H_0 \equiv H_S + H_B$ defines a standard interaction picture in which the coupling $\tilde{V}(t) = e^{\im H_0 t/\hbar} V(t) e^{-\im H_0 t/\hbar} $ determines the evolution of the system-bath density operator $\tilde{w}(t)$.  To second order in the coupling its instantaneous rate of change is
\begin{equation}
\dot{\tilde{w}} = -\frac{\im}{\hbar} [\tilde{V}(t),\tilde{w}(0)] -  \frac{1}{\hbar^2}\int_0^t dt'\,  [\tilde{V}(t),[\tilde{V}(t'),\tilde{w}(0)]].
\end{equation}

Next, we assume the bath is in a thermal state $\rho_B =
(e^{-H_B/k_BT})/Z_B$ initially uncorrelated with the chain:
$\tilde{w}(0) = \rho(0) \otimes\rho_B$.  Here $T$ is the temperature
and $Z_B=\sum_n e^{-E_n/k_BT}$ is the bath's partition function.  We
average (trace) over the states of the bath to obtain a preliminary
equation of motion for the system:
\begin{equation}
\dot{\tilde{\rho}} \!=\!\! -\frac{\lambda^2}{\hbar^2} \!\! \sum_{j,k} \! \int_0^t \!\!\!\! dt'\, \mbox{Tr}_B[\tilde{{\cal{X}}}_j(t)\tilde{\phi}_j(t),[\tilde{{\cal{X}}}_k(t')\tilde{\phi}_k(t'),\tilde{w}(0)]]
\label{prelim0}
\end{equation}
(the first order term vanished because $\mbox{Tr}_B[\tilde{\phi}_j \, \rho_B] =0$).

\begin{figure}
\begin{center}
\includegraphics{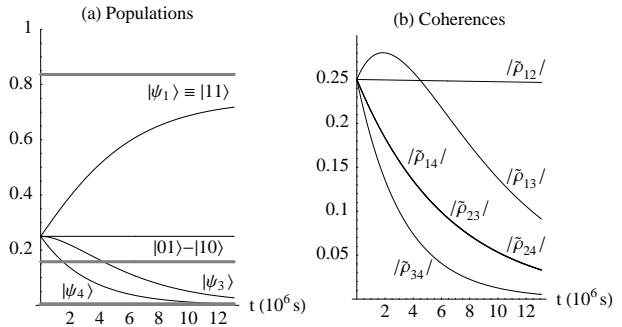}
\end{center}
\caption{Dicke Super- and Sub-radiance and Super- and Sub-decoherence in two small SQUIDs dissipating collectively.  $\ket{\psi_1}+\ket{\psi_2}+\ket{\psi_3}+\ket{\psi_4}$ cooling above the critical point.  (a) $\ket{\psi_4}$ flows into $\ket{\psi_3}$ which flows into $\ket{\psi_1}$.  The singlet $\ket{\psi_2}=\ket{01}-\ket{10}$ is a protected subspace and cannot relax into $\ket{\psi_1}$; the populations do not approach their thermal levels (gray-lines).  But the collective relaxation rate from $\ket{\psi_3}$ into $\ket{\psi_1}$ is double the independent rate.  (b) The coherence $\tilde{\rho}_{12}$, between $\ket{\psi_1}$ and $\ket{\psi_2}$, barely decays; the only mechanism for that is the $\ket{\psi_1} \rightarrow \ket{\psi_3}$ transition which is suppressed by the cold bath.  Note also how $\tilde{\rho}_{34}$ flows, in a pair of resonant transitions, into $\tilde{\rho}_{13}$ whose collective decoherence rate is double its independent dissipation rate.  ($8J/h = 1.0$GHz, $2B/h=1.5$GHz, $k_BT/h=0.3$GHz, $R=10 $nm, $I=0.1 \mu$A).}
\label{2SQUIDsCollective}
\end{figure}

For a given $(j,k)$-pair of SQUIDs, the trace of the nested commutators results in four integrals.  They differ only in the permutation of terms; a representative one is
\begin{equation}
\tilde{{\cal{X}}}_j(t)\underbrace{\int_0^t dt' \,\tilde{{\cal{X}}}_k(t') \overbrace{\mbox{Tr}_B[\tilde{\phi}_j(t)\tilde{\phi}_k(t')\rho_B]}^{\nu_{jk}(\tau)-\im\mu_{jk}(\tau)}}_{\tilde{\cal{V}}_{jk}(t)-\im\,\tilde{\cal{U}}_{jk}(t)} \rho(0) ,
\label{averaging0}
\end{equation}
where the over- and under-braces highlight the time-averaging of the
interaction operator $\tilde{\cal{X}}_k(t')$, into what we call the
noise $\tilde{\cal{V}}_{jk}(t)$ and susceptibility
$\tilde{\cal{U}}_{jk}(t)$ operators, weighted by the kernels
$\nu_{jk}(\tau)$ and $\mu_{jk}(\tau)$ that, with $\tau \equiv t-t'$,
are the real and imaginary parts of the bath correlator $\mbox{Tr}_B
[\tilde{\phi}_j(t) \, \tilde{\phi}_k(t') \, \rho_B]$. (In the context
of quantum Brownian oscillator systems, the susceptibility kernel
used here is called dissipation $\eta$ \cite{HPZ92} there.)

The four integrals for each SQUID-pair, originating from the nested commutators and each contributing a noise and a susceptibility, can be collected to obtain a ``Born'' (but not yet Born-Markov) master equation (still in the interaction picture, still only valid to second order),
\begin{equation}
\dot{\tilde{\rho}} = -\frac{\lambda^2}{\hbar^2}\sum_{j,k=1}^N
\left(
[\tilde{\cal{X}}_j,[\tilde{\cal{V}}_{jk},\tilde{\rho}]] - \im [\tilde{\cal{X}}_j,\{\tilde{\cal{U}}_{jk},\tilde{\rho}\}]
\right)
,\label{Born0}
\end{equation}
where we have used Born's approximation that replaces $\rho(0)$ with $\tilde{\rho}(t)\approx\rho(0)$.

Born's approximation adds an updating to the integrated solution
for $\tilde{\rho}(t)$; in a small time step $dt$ the instantaneous
change $d\tilde{\rho}$ depends not on the initial $\rho(0)$ but on
the updated instantaneous $\tilde{\rho}(t)$.  We will see that this
updating is needed for the long-time equilibration to thermal
equilibrium.  The Born master equation, although technically still
only valid to second order, is a plausible guess at the longer-time
open system dynamics which, like any other theory, can only be
supported by real data or exactly solvable open systems, like the
quantum Brownian oscillator, and may not be valid in every case.

\begin{figure}
\begin{center}
\includegraphics{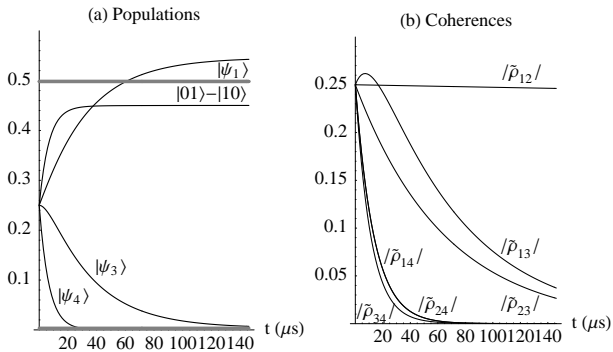}
\end{center}
\caption{Critical behavior of two large SQUIDs.  $\ket{\psi_1}+\ket{\psi_2}+\ket{\psi_3}+\ket{\psi_4}$ cooling at the critical point.  The distant (e.g.\ $40\mu$m) SQUIDs are closely-spaced compared to the bath wavelengths corresponding to the frequency $\omega_{12}=0$;  the $\ket{\psi_1} \leftrightarrow \ket{\psi_2}$ transitions are blocked, akin to the collective behavior of two small SQUIDs close together.  (a) $\ket{\psi_4}$ flows into $\ket{\psi_3}$ and $\ket{\psi_2}=\ket{01}-\ket{10}$, whereas $\ket{\psi_3}$ flows only into $\ket{\psi_1}$, giving it an excess population over $\ket{\psi_2}$ which the chain cannot quickly resolve; the $\ket{\psi_2}\rightarrow\ket{\psi_4}\rightarrow\ket{\psi_3}\rightarrow\ket{\psi_1}$ pathway takes an extraordinarily long time (not shown).  (b) The coherence $\tilde{\rho}_{12}$ barely decays; transitions $\ket{\psi_2}\leftrightarrow\ket{\psi_1}$ are blocked while transitions from $\ket{\psi_1}$ and $\ket{\psi_2}$ are suppressed by the cold bath.  Some of the decaying $\tilde{\rho}_{34}$ is absorbed by $\tilde{\rho}_{13}$.  ($8J/h = 1.0$GHz, $2B/h=1.0$GHz, $k_BT/h=0.2$GHz, $R=10 \mu$m, $I=3 \mu$A).}
\label{2SQUIDsCritical}
\end{figure}

\subsection{The Markov Approximation}
Next, we time-average the interaction operators $\tilde{\cal{X}}_k(t')$ into the noise $\tilde{\cal{V}}_{jk}(t)$ and susceptibility $\tilde{\cal{U}}_{jk}(t)$ operators by making a Fourier expansion of the interaction operators using amplitudes $\tilde{\cal{X}}_k^{\omega}(t)$ and $\tilde{\cal{P}}_k^{\omega}(t)$:
\begin{equation}
\tilde{\cal{X}}_k(t-\tau) = \sum_{\omega\geq0}  \left(\tilde{\cal{X}}_k^{\omega}(t) \cos(\omega \tau) -  \tilde{\cal{P}}_k^{\omega}(t) \sin(\omega \tau)\right)\!.
\label{XandP}
\end{equation}
(In the eigenbasis of $H_S$, the non-zero matrix elements of $\tilde{\cal{X}}_k^{\omega}$ are just those matrix elements of $\tilde{\cal{X}}_k$ with energy difference $\pm \hbar \omega$; multiplying those same matrix elements by $\mp \im$ gives $\tilde{\cal{P}}_k^{\omega}$.  (This sign convention is consistent with the matrix elements of a harmonic oscillator's position, $\bra{n}a^{\dagger}+a\ket{n+1} = \sqrt{n+1}$, and momentum, $\im \bra{n}a^{\dagger}-a\ket{n+1} = -\im\sqrt{n+1}$.)  $\tilde{\cal{X}}_k^0$ consists primarily of the diagonal matrix elements of $\tilde{\cal{X}}_k$ but also includes matrix elements between degenerate states.)

We thereby obtain basis-independent expressions for the noise,
\begin{equation}
\tilde{\cal{V}}_{jk}(t) = \sum_{\omega\geq0} \left(D_{jk}^{\omega}(t)\tilde{\cal{X}}_k^{\omega}(t)   - A_{jk}^{\omega}(t) \tilde{\cal{P}}_k^{\omega}(t)  \right),
\end{equation}
and susceptibility,
\begin{equation}
\tilde{\cal{U}}_{jk}(t) = \sum_{\omega\geq0} \left(r_{jk}^{\omega}(t)\tilde{\cal{X}}_k^{\omega}(t)   -  \gamma_{jk}^{\omega}(t) \tilde{\cal{P}}_k^{\omega}(t) \right),
\end{equation}
where the coefficients (to be discussed shortly) of diffusion $D$,
anomalous diffusion $A$, renormalization $r$, and damping $\gamma$
serve to Fourier-sample (at least for $t\rightarrow\infty$) the real
($\nu_{jk}$) and imaginary ($\mu_{jk}$) parts of the bath correlator:
\begin{equation}
\begin{array}{ccc}
D_{jk}^{\omega}(t) \!=\! \int_0^t \! d\tau \nu_{jk}(\tau) \cos\omega\tau & & A_{jk}^{\omega}(t) \!=\! \int_0^t \! d\tau \nu_{jk}(\tau) \sin\omega\tau \\
r_{jk}^{\omega}(t) \!=\! \int_0^t \! d\tau \mu_{jk}(\tau) \cos\omega\tau & &  \gamma_{jk}^{\omega}(t) \!=\! \int_0^t \! d\tau \mu_{jk}(\tau) \sin\omega\tau.
\end{array}
\end{equation}

Finally, we make the Markov approximation, that uses constant
coefficients obtained in the limit $t \rightarrow \infty$ in place of
the time-dependent ones, and substitute these noise- and
susceptibilty-operators into Eq. \eqn{Born0} to obtain our
Born-Markov master equation,
\[
\dot{\tilde{\rho}} =\frac{\lambda^2}{\hbar^2} \sum_{j,k=1}^N \sum_{\omega \geq 0}
\mbox{\Large(} \im \, r_{jk}^{\omega} [\tilde{\cal{X}}_j,\{\tilde{\cal{X}}_k^{\omega},\tilde{\rho}\}]  - \im \, \gamma_{jk}^{\omega} [\tilde{\cal{X}}_j,\{\tilde{\cal{P}}_k^{\omega},\tilde{\rho}\}]\]
\vspace{-0.25in}
\begin{equation}
 \;\;\;\;\;\;\;\;\;\;\; \;\;\;\;\;\;\;\;\; - D_{jk}^{\omega} [\tilde{\cal{X}}_j,[\tilde{\cal{X}}_k^{\omega},\tilde{\rho}]] +A_{jk}^{\omega} [\tilde{\cal{X}}_j,[\tilde{\cal{P}}_k^{\omega},\tilde{\rho}]]
\mbox{\Large)}.
\label{BornMarkov0}
\end{equation}
This is essentially a Fourier-series version of the Born master equation using Markov (constant) coefficients.  It is reassuring that, in the limit of only one contact point (no sum over $j,k$) and only one energy splitting (no sum over $\omega$) it becomes
\begin{equation}
\dot{\tilde{\rho}} \!\sim\!
\im  r [\tilde{x},\{\tilde{x},\tilde{\rho}\}]  \!-\! \im \lambda [\tilde{x},\{\tilde{p},\tilde{\rho}\}] \!-\! D [\tilde{x},[\tilde{x},\tilde{\rho}]] \!+\!A [\tilde{x},[\tilde{p},\tilde{\rho}]]
\end{equation}
which is the well-known \cite{Primer, HPZ92} Born-Markov equation for the quantum Brownian motion of an oscillator system ($H_S= \hbar \omega a^{\dagger} a$) with a bilinear coupling ($V\sim x \phi$) to an oscillator bath.

\subsection{Cross Term Coefficients and Spectral Densities}
In appendices~\ref{correlator}, \ref{spectraldensities}, and \ref{constantcoefficients} we study the bath correlator and obtain the coefficients of diffusion
\begin{equation}
D^{\omega}_{jk}  \equiv \frac{\pi}{2} J_{jk}(\omega) \coth(\frac{\hbar\omega}{2 k_BT}) = \lim_{t\rightarrow\infty} D^{\omega}_{jk}(t)
\end{equation}
and damping
\begin{equation}
\gamma^{\omega}_{jk}  \equiv \frac{\pi}{2} J_{jk}(\omega) = \lim_{t\rightarrow\infty} \gamma^{\omega}_{jk}(t),
\end{equation}
written in terms of spectral densities $J_{jk}(\omega)$ with, e.g.,
\begin{equation}
J_{jj}(\omega) = \frac{\hbar m_e^2 I^2 R^4 \omega^5}{6 \lambda^2 e^2 \rho c_{\perp}^5} {}_pF_q(\{\frac{3}{2}\},\!\{\frac{5}{2},3\},\!-(\omega \tau_R)^2)
\end{equation}
the individual spectral density for a single ($j=k$) SQUID.  Here ${}_pF_q$ is the generalized hypergeometric function and $\tau_R \equiv R/c_{\perp}$ is half the time it takes a phonon to traverse the SQUID.

We then find the {\em same} spectral density to first order in small $\vec{k}\cdot\vec{r}_{jk} = \omega\,\tau_{jk} \ll 1$ for a distinct ($j\neq k$) pair of SQUIDs separated by $\vec{r}_{jk}$, where $\vec{k}$ is a bath wavevector at the chain transition frequency $\omega$ and $\tau_{jk} = r_{jk}/c_{\perp}$ is the phonon transit time between the SQUIDs.   Remarkably, even ``large'' and ``far-apart'' SQUIDs can have $\omega\tau_{jk}\ll1$ at a level crossing where $\omega \approx 0$.  In these cases, as for close-together small SQUIDs, we thus use
\begin{equation}
J_{jk}(\omega) = J_{jj}(\omega) {\mbox{ when }} \omega\,\tau_{jk} \ll 1.
\end{equation}

On the other hand, the cross term spectral densities vanish for truly far-apart SQUIDs satisfying $\vec{k}\cdot\vec{r}_{jk} = \omega\,\tau_{jk} \gg1$:
\begin{equation}
J_{jk}(\omega) = 0 {\mbox{ when }} \omega\,\tau_{jk} \gg 1.
\end{equation}

We avoid the intermediate regime by tuning the chain Hamiltonian close-to or far-from any critical point energy-level crossings.

In our Born-Markov master equation ~\eqn{BornMarkov0}, the sum over
transition frequencies thus splits into two parts, one for $\omega
\tau_{jk} \gg 1$, for which all cross terms vanish, and one for
$\omega \tau_{jk} \ll 1$, for which the cross terms are equally
weighted.  For example, the diffusion term becomes
\begin{equation}
\sum_{j,k}\sum_{\omega \gg 1/\tau_{jk}} \!\!\!\!\!  \delta_{jk} \, D_{jj}^{\omega} [\tilde{\cal{X}}_j,[\tilde{\cal{X}}_k^{\omega},\tilde{\rho}]] + \sum_{j,k} \sum_{\omega  \ll 1/\tau_{jk}} \!\!\!\!\! D_{jj}^{\omega} [\tilde{\cal{X}}_j,[\tilde{\cal{X}}_k^{\omega},\tilde{\rho}]].
\end{equation}

The case $\omega \tau_{jk} \ll1$ is the provenance of a collective coupling to the bath $V = \lambda (\sum_j {\cal{X}}_j) \phi_1$ (whose square includes all cross terms equally).  In our case, $\sum_j {\cal{X}}_j = \sum_j \sigma_j^z \equiv J_z$.  A collective coupling can have degenerate subspaces with which the system Hamiltonian may (or may not) cooperate, giving a decoherence- and/or relaxation-free subspace (or not) \cite{Zanardi, DuanGuo,LidarDFS}.  On the other hand, when we include all the cross terms equally, the master equation has $N$ times as many terms, for $N$ SQUIDs, as in the case of independent dissipation.  This can scale the decoherence and relaxation rates linearly with the number of subsystems \cite{DickeSuper, ChudnovskySuper}.

\begin{figure}
\begin{center}
\includegraphics{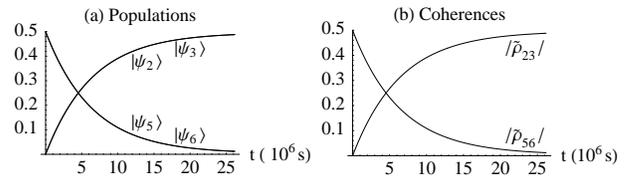}
\end{center}
\caption{Automatic quantum error correction \cite{BarnesWarren} in three small SQUIDs dissipating collectively.  $\ket{\psi_5}+\ket{\psi_6}$ cooling to $\ket{\psi_2}+\ket{\psi_3}$.  (a) The collective interaction operator $J_z={\cal{X}}_1+{\cal{X}}_2+{\cal{X}}_3$ breaks the network of allowed transitions into three isolated pieces:  $\ket{\psi_5}\leftrightarrow\ket{\psi_2}$; $\ket{\psi_6}\leftrightarrow\ket{\psi_3}$; and $\ket{\psi_8}\leftrightarrow\ket{\psi_7}\leftrightarrow\ket{\psi_4}\leftrightarrow\ket{\psi_1}$.  (b)  The $\tilde{\rho}_{56}$ coherence flows along with the populations into $\tilde{\rho}_{23}$ because $\omega_{25}=\omega_{36}$.  Then $\tilde{\rho}_{23}$ never decays; transitions back into $\ket{\psi_5}$ and $\ket{\psi_6}$ are suppressed by the cold bath and are resonant (and therefore coherent) anyway.  ($6J/h = 1.0$GHz, $2B/h=1.5$GHz, $k_BT/h=0.3$GHz, $R=10 $nm, $I=0.1 \mu$A).}
\label{3SQUIDsCollective}
\end{figure}

\section{The Four Generators}
\subsection{The Renormalization and Anomalous Diffusion}
The master equation has four generators, one for each of the coefficients.  It can be shown \cite{SkinnerPhD}, that for a weak coupling of our Heisenberg SQUID chain to the bath and with $J$ and $B$ set close-to or far-from any resonant pairs of allowed transitions, the renormalization and anomalous diffusion contribute effectively Hamiltonian dynamics, which can be dropped from the master equation by ``renormalizing'' the chain Hamiltonian.

We would like to concentrate on the damping and diffusion.  To make
concrete sense of these two generators, our examples use the familiar
and useful energies and eigenstates of the Heisenberg SQUID Chain.
In other words, we assume that the renormalized Hamiltonian is given
by Eq. \eqn{H_S} as though the ``bare'' SQUIDs were engineered with a
slightly different original Hamiltonian, including a ``counter term''
that cancels the renormalization and anomalous diffusion caused by
their crystal environment.  This engineering may actually be quite
difficult, relying perhaps on many trials and errors.  In any case,
our general discussion of the damping and diffusion is in terms of
the renormalized energies and the matrix elements in the renormalized
eigenbasis.

\subsection{The Damping and Diffusion}
We now show that the damping and diffusion work together to
effectively decohere and thermalize the system within each network of
allowed transitions.  In Eq. \eqn{BornMarkov0} we set
$r_{jk}^{\omega}=0=A_{jk}^{\omega}$ in accordance with the
aforementioned renormalization and obtain
\begin{equation}
\dot{\tilde{\rho}} \!=\! -\frac{\lambda^2}{\hbar^2}\!\sum_{j,k,\omega}
\!\!\left(
D_{jk}^{\omega}[\tilde{\cal{X}}_j,[\tilde{\cal{X}}_k^{\omega},\tilde{\rho}]] \!+\! \im \gamma_{jk}^{\omega} [\tilde{\cal{X}}_j,\{\tilde{\cal{P}}_{jk}^{\omega},\tilde{\rho}\}]
\right)\!.
\end{equation}

We expand out the commutators, work in the energy eigenbasis (where $\hbar \omega_{\alpha\beta} = E_{\beta}-E_{\alpha}$ is the energy lost in the transition, at frequency $\omega_{\alpha\beta}$, from $\ket{\beta}$ to $\ket{\alpha}$) by inserting resolutions of the identity, e.g.\ $I=\sum_{\beta}\ket{\beta}\bra{\beta}$, and define the rates
\begin{equation}
\Gamma_{jk}^{\alpha\beta} \equiv \frac{\lambda^2}{\hbar^2}  \underbrace{\frac{\pi}{2} J_{jk}(\omega_{\alpha\beta})\left(1+ \coth (\frac{\hbar \omega_{\alpha\beta}}{2 k_B T})\right)}_{\gamma_{jk}^{\omega_{\alpha\beta}}+D_{jk}^{\omega_{\alpha\beta}}}
\end{equation}
to obtain a matrix element equation, in terms of the populations $\tilde{\rho}_{\alpha\alpha} \equiv \bra{\alpha}\tilde{\rho}\ket{\alpha}$ and coherences $\tilde{\rho}_{\alpha\delta} \equiv \bra{\alpha}\tilde{\rho}\ket{\delta}$, for their rates of change,
\[
\dot{\tilde{\rho}}_{\alpha\delta} = - \sum_{j,k}
\mbox{\Large(}
\sum_{\beta\bar{\alpha}} \tilde{\cal{X}}_{j\alpha\beta}\tilde{\cal{X}}_{k\beta\bar{\alpha}}\Gamma_{jk}^{\beta\bar{\alpha}} \tilde{\rho}_{\bar{\alpha}\delta}
-\sum_{\beta\gamma}
\tilde{\cal{X}}_{j\alpha\beta}\tilde{\cal{X}}_{k\gamma\delta} \Gamma_{jk}^{\delta\gamma} \tilde{\rho}_{\beta\gamma}
\]
\vspace{-0.25in}
\begin{equation}
-\sum_{\beta\gamma}
\tilde{\cal{X}}_{k\alpha\beta}\tilde{\cal{X}}_{j\gamma\delta} \Gamma_{jk}^{\alpha\beta} \tilde{\rho}_{\beta\gamma}
+\sum_{\bar{\delta}\beta} \tilde{\cal{X}}_{k\bar{\delta}\beta} \tilde{\cal{X}}_{j\beta\delta}\Gamma_{jk}^{\beta\bar{\delta}} \tilde{\rho}_{\alpha\bar{\delta}}
\mbox{\Large)}
,\end{equation}
where we have used the evenness of $D_{jk}^{\omega}$ and oddness of $\gamma_{jk}^{\omega}$ with respect to $\omega$ to combine the two generators into this one expression.

Many terms in this sum are suppressed by selection-ruled resonance conditions.  In the second term, for example, most of the $\tilde{\cal{X}}_{j\alpha\beta}\tilde{\cal{X}}_{k\gamma\delta} \sim e^{-\im(\omega_{\alpha\beta}+\omega_{\gamma\delta})t}$ oscillate so quickly, compared to the weak coupling between the chain and the bath, that they average to zero unless there is a near-resonance $\omega_{\alpha\beta} \approx \omega_{\delta\gamma}$.  The second term is thus effectively a sum over nearly-resonant pairs of allowed transitions from states $\ket{\beta}$ and $\ket{\gamma}$ into $\ket{\alpha}$ and $\ket{\delta}$ respectively.  So is the third term.

\begin{figure}
\begin{center}
\includegraphics{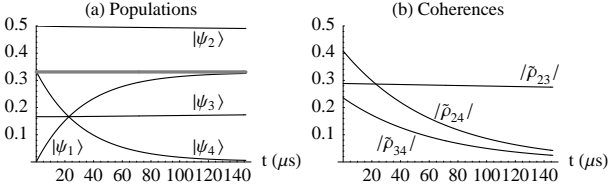}
\end{center}
\caption{Critical behavior of three large SQUIDs dissipating independently.  $\ket{\uparrow\uparrow\downarrow}$ cooling at the critical point.  The distant SQUIDs are closely-spaced compared to the bath wavelengths corresponding to the frequencies $\omega_{12}=\omega_{13}=\omega_{23}=0$ and $\omega_{45}=\omega_{46}=\omega_{56}=0$.  (a)  The $\ket{\psi_2}  \leftrightarrow \ket{\psi_1}$ and $\ket{\psi_3} \leftrightarrow \ket{\psi_1}$ transitions are cut out of the network while upward transitions are suppressed by the cold bath; therefore $\ket{\psi_2}$ and $\ket{\psi_3}$ have long-lived populations.  (b)  $\tilde{\rho}_{23}$ is long-lived because the transitions out of its $\ket{\psi_2}$ and $\ket{\psi_3}$ supports are so rare.  ($6J/h = 1.0$GHz, $2B/h=1.0$GHz, $k_BT/h=0.1$GHz, $R=10 \mu$m, $I=3 \mu$A).}
\label{3SQUIDsCritical}
\end{figure}

Meanwhile, the first and fourth terms are effectively sums over allowed transitions from states $\ket{\bar{\alpha}}$ near $E_{\alpha}$ ($\omega_{\beta\bar{\alpha}} \approx \omega_{\beta\alpha}$) and states $\ket{\bar{\delta}}$ near $E_{\delta}$ ($\omega_{\beta\bar{\delta}} \approx \omega_{\beta\delta}$) into those states $\ket{\beta}$ that are accessible by transitions from $\ket{\alpha}$ and $\ket{\delta}$ respectively.

In the first and fourth terms, the remaining transitions from states near $E_{\alpha}$ ($\omega_{\beta\bar{\alpha}} \approx \omega_{\beta\alpha}$) and $E_{\delta}$ ($\omega_{\beta\bar{\delta}} \approx \omega_{\beta\delta}$) tend to decrease $\tilde{\rho}_{\alpha\delta}$ in proportion to $\tilde{\rho}_{\alpha\delta}$ as well as in proportion to nearby (in energy) matrix elements $\tilde{\rho}_{\bar{\alpha}\delta}$ (in the same column) and  $\tilde{\rho}_{\alpha\bar{\delta}}$ (in the same row).  In the second and third terms, the remaining nearly-resonant pairs of transitions from $\ket{\beta}$ and $\ket{\gamma}$ into $\ket{\alpha}$ and $\ket{\delta}$ ($\omega_{\alpha\beta} \approx \omega_{\delta\gamma}$) tend to increase $\tilde{\rho}_{\alpha\delta}$ in proportion to those matrix elements $\tilde{\rho}_{\beta\gamma}$ within, and/or nearly within, the diagonal that includes $\tilde{\rho}_{\alpha\delta}$ (when the density matrix is stretched to be linearly spaced with increasing energy).  When $\delta=\alpha$ this diagonal is the central diagonal, sometimes called\ {\em{the}} diagonal.

\subsection{Decoherence and Relaxation}
Two general features of these coupled first order differential equations are decoherence and relaxation.  Decoherence is caused by transitions from $\ket{\alpha}$ and $\ket{\delta}$,
\[
\frac{\partial \dot{\tilde{\rho}}_{\alpha\delta}}{\partial \tilde{\rho}_{\alpha\delta}} = -
\sum_{j,k}
\mbox{\Large(}
\sum_{\beta} \tilde{\cal{X}}_{j\alpha\beta}\tilde{\cal{X}}_{k\beta\alpha}\Gamma_{jk}^{\beta\alpha}
-\overbrace{\tilde{\cal{X}}_{j\alpha\alpha}\tilde{\cal{X}}_{k\delta\delta} \Gamma_{jk}^{\delta\delta}}^{0}
\]
\vspace{-0.25in}
\begin{equation}
-\underbrace{\tilde{\cal{X}}_{k\alpha\alpha}\tilde{\cal{X}}_{j\delta\delta} \Gamma_{jk}^{\alpha\alpha}}_{0}
+\sum_{\beta} \tilde{\cal{X}}_{k\delta\beta} \tilde{\cal{X}}_{j\beta\delta}\Gamma_{jk}^{\beta\delta} \mbox{\Large)}  \! \equiv \!  -\bar{\Gamma}_{\alpha\delta}
\end{equation}
(here the second and third terms vanished because of our specific selection rules), and is exacerbated by transitions from nearby states.  Setting $\delta=\alpha$ we see the relaxation dynamics, in that the population $\tilde{\rho}_{\alpha\alpha}$ is flowing to and from the $\tilde{\rho}_{\beta\beta}$ at the selection-ruled transition rates $2  \sum_{jk} {\cal{X}}_{j\alpha\beta}{\cal{X}}_{k\beta\alpha}\Gamma_{jk}^{\beta\alpha} \tilde{\rho}_{\alpha\alpha}$ and
$2 \sum_{jk} {\cal{X}}_{j\alpha\beta}{\cal{X}}_{k\beta\alpha}\Gamma_{jk}^{\alpha\beta} \tilde{\rho}_{\beta\beta}$.  A stationary (and thermal) balance is eventually reached at $\tilde{\rho}_{\beta\beta}/\tilde{\rho}_{\alpha\alpha} = \Gamma_{jk}^{\beta\alpha}/\Gamma_{jk}^{\alpha\beta} = e^{-\hbar\omega_{\alpha\beta}/k_BT}$ (at which point the populations' effect on the off-diagonal coherences $\dot{\tilde{\rho}}_{\alpha\delta}$,
\begin{equation}
-\!\!
\sum_{j,k,\beta}
\!\tilde{\cal{X}}_{j\alpha\beta} \tilde{\cal{X}}_{k\beta\delta}
\mbox{\Large(}
\underbrace{\Gamma_{jk}^{\beta\delta} \tilde{\rho}_{\delta\delta}}_{\bar{\alpha}=\delta} - \underbrace{\Gamma_{jk}^{\delta\beta} \tilde{\rho}_{\beta\beta}}_{\gamma=\beta}
-\underbrace{\Gamma_{jk}^{\alpha\beta}\tilde{\rho}_{\beta\beta}}_{\gamma=\beta} +\underbrace{\Gamma_{jk}^{\beta\alpha}\tilde{\rho}_{\alpha\alpha}}_{\bar{\delta}=\alpha}
\!\mbox{\Large)}
,
\end{equation} also vanishes).  The decay of the coherences allows the relaxation to proceed to a thermal equilibrium $\rho_T \equiv e^{-H_S/k_BT}/Z_S$ provided the network of selection-ruled transitions does not isolate any subspace(s).  For isolated networks each subspace will obtain its own a stationary thermal balance of populations constrained by the total available initial probability to be in that subspace.


Remarkably, the decay of $\tilde{\rho}_{\alpha\delta}$ is offset, as it is on the central diagonal, by the nearly-resonant pairs of transitions into $\ket{\alpha}$ and $\ket{\delta}$ from any matrix elements $\tilde{\rho}_{\beta\gamma}$ within, and/or nearly within, the off-center diagonal that includes $\tilde{\rho}_{\alpha\delta}$.  Indeed, a population-like coherence flow is established between $\tilde{\rho}_{\alpha\delta}$ and $\tilde{\rho}_{\beta\gamma}$ that is primarily into the lower energy subspace when the temperature is low; quantum oscillations which are decaying in one subspace can in principle emerge in a lower energy subspace.

\section{Numerical Simulations}
We simulate the effects of the matrix element equation in a variety of scenarios for two- and three-SQUID chains.  In all cases we choose $J$ and $B$ to set the chain close-to or far-from resonant pairs of transitions (we can run at the critical point because of its exactly-resonant pairs).  That way we can and do discard the oscillating coefficients from the equation, since their effect would average to zero anyway.  The matrix element equation becomes a coupled first order differential equation with constant coefficients which we numerically integrate \cite{Mathematica}.

We consider large SQUIDs, $R=10 \mu$m with $I=3\mu$A, and small SQUIDs, $R=10$nm with $I=0.1\mu$A.  In both cases we imagine them to be spaced $4R$ apart in a solid crystal with mass density $\rho=5$g$/$cm${}^3$ and sound velocity $5$km$/$s.  We set the Heisenberg splitting at $1.0$GHz $=8J/h$ for a two-SQUID chain and $1.0$GHz $=6J/h$ for a three-SQUID chain.  We then choose SQUIDs with frequencies of $2B/h = 0.5$, $1.0$, or $1.5$ GHz (below, at, or above the chain's critical point).  We set the temperature to be $1/5$ the SQUID frequencies, i.e.\ $k_BT/h=0.1$, $0.2$, or $0.3$ GHz.

Although larger SQUIDs are possible, these parameters keep the photon-induced decoherence rates \cite{Chudnovsky2}, in the absence of shielding, well below our phonon-induced rates.  But still these SQUIDs' $4R$ separation is large enough to discard the $j\neq k$ cross terms.  At the critical point there are certain transition frequencies $\omega=0 \ll 1/\tau_{jk}$ whose cross term rates $\Gamma_{jk}^{\omega=0}$ are identical to the $j=k$ rates, giving a partially collective coupling to the bath.  On the other hand, the small SQUIDs' $4R$ separation is small enough to achieve a collective coupling to the bath for all $\omega$.

In either case the matrix elements of the interaction operators are used to calculate all the constant coefficients in the matrix element equation and we then proceed with the numerical simulations for any initial state $\rho(0)$.

\section{Discussion}
The relaxation and decoherence will thermalize the SQUIDs when the network of selection-rules is complete.  For example, for two large SQUIDs dissipating independently, the matrix elements of the interaction operators show that the network of selection-rules is $\ket{\psi_1} \leftrightarrow \ket{\psi_2} \leftrightarrow \ket{\psi_4}$ and $\ket{\psi_1} \leftrightarrow \ket{\psi_3} \leftrightarrow \ket{\psi_4}$.  There is a reasonable pathway from any eigenstate to any other and these examples lead to thermalization.  However, in the case of small SQUIDs close together there is a collective coupling $J_z$ and the network (for two SQUIDs) becomes $\ket{\psi_1} \leftrightarrow \ket{\psi_3} \leftrightarrow \ket{\psi_4}$ while $\ket{\psi_2}$ is a protected subspace.  We attempt to summarize with a schematic:
\[
\begin{array} {ccccccccccccccccccccccc}
& & 4 & & & & & & & & 4 & & & & & & & & 4 & &  \\
& \nearrow\!\!\!\!\!\!\swarrow & & \searrow\!\!\!\!\!\!\nwarrow & & & & & & \nearrow\!\!\!\!\!\!\swarrow & & {\;\;}  & & & & & & \nearrow\!\!\!\!\!\!\swarrow & & \searrow\!\!\!\!\!\!\nwarrow & \\
3 & & \mbox{ind} & & 2 & & & & 3 & & \mbox{col} & & 2 & & & & 3 & & \mbox{cri} & & 2  \\
& \searrow\!\!\!\!\!\!\nwarrow & &  \nearrow\!\!\!\!\!\!\swarrow & & & & & & \searrow\!\!\!\!\!\!\nwarrow & &  {\;\;} & & & & & & \searrow\!\!\!\!\!\!\nwarrow & &  {} & \\
& & 1 & & & & & & & & 1 & & & & & & & & 1 & &  \\
\end{array}
\]
in which the first and second networks are those of independent and collective dissipation.  The third network is at the critical point $\omega_{12}=0$ for which the network is $\ket{\psi_1}\leftrightarrow\ket{\psi_3}\leftrightarrow\ket{\psi_4}\leftrightarrow\ket{\psi_2}$.  Being at the critical point severs the $\ket{\psi_1}\leftrightarrow\ket{\psi_2}$ link, as was done in the collective case, for $\omega_{12}\tau_{jk} =0 \ll 1$, but not the $\ket{\psi_2}\leftrightarrow\ket{\psi_4}$ link for which $\omega_{14}\tau_{jk}\gg1$.  The only allowed transitions out of $\ket{\psi_1}$ and $\ket{\psi_2}$ are suppressed by the cold bath which is loath to supply the necessary energy.  This helps to protect the population of $\ket{\psi_2}$ and the coherence $\tilde{\rho}_{12}$.

\section{Conclusion}
A chain of a few coupled SQUIDs exchanging their angular momenta with a phonon bath can be studied, in the Born-Markov approximation, with master equation techniques from the quantum Brownian motion model.  The damping and diffusion give a matrix element equation showing decoherence, relaxation, and the possibility for decaying quantum oscillations to emerge in a lower energy subspace.  The relaxation adjusts the populations of the eigenstates and undermines their support for any superposition (coherence) between them, leading to decoherence.

The cascade of populations can occupy entangled states of intermediate energy, resulting in a surge of the entanglement of formation that indicates the number of singlets needed to form, from local operations and classical communication, an ensemble of SQUID pairs $\rho(t)$ \cite{Wootters}.  The entanglement is induced even though the SQUIDs are dissipating independently.

The level spacings in the Heisenberg SQUID chain include pairs of resonant transitions which are necessary for coherent oscillations to decay into a lower energy subspace where they can decohere more slowly.  In this phenomenon, a superposition of two eigenstates relaxes coherently into a superposition of two lower-energy eigenstates with the same energy difference as the upper two.

Small SQUIDs close together exhibit a collective coupling to the bath which can give a protected subspace and enhanced or suppressed transition and decoherence rates.  In effect, the network of selection-ruled transitions is broken into isolated pieces.  When the level spacings cooperate to allow coherence flow in these sufficiently isolated pieces, decaying quantum information can reappear and be sustained in a lower energy subspace; this is the idea behind ``automatic quantum error correction'' \cite{BarnesWarren}.

Another feature of the Heisenberg SQUID chain is the critical point
level crossings where an allowed transition vanishes along with its
frequency.  The network of selection-rules in effect acquires some
features of the collective behavior as $\omega\tau_{jk} \rightarrow 0
\ll 1$;  even large SQUIDs spaced well apart, when tuned to the
critical point, can have extended coherence times.  More generally,
this suggests some qubit encoding schemes might be augmented with
inter-qubit couplings to obtain some or all of the benefits of a
collective coupling to the bath.\\

{\bf Acknowledgment} This work is supported in part by grants from
the NSF-ITR program (PHY-0426696), NIST and NSA-LPS. We are grateful
for helpful discussions with E.\ Chudnovsky, F.\ Strauch, P.\
Johnson, and S.\ Shresta.

\appendix
\section{Coherences and Populations\label{cohpop}}
The instantaneous state of a spin-chain can be described by a density operator $\rho(t) = \sum_n p_n \ket{\psi_n}\bra{\psi_n}$ which averages ``outer products'' of pure states $\ket{\psi_n}\bra{\psi_n}$, weighted by their probabilities $p_n$, into a statistical mixture.  In the energy eigenbasis, a diagonal element $\bra{\alpha}\rho\ket{\alpha}$ is the probability of obtaining the eigenstate $\ket{\alpha}$, sometimes called the population of $\ket{\alpha}$.  An off-diagonal element $\bra{\alpha}\rho\ket{\beta}$ results from including superpositions of eigenstates, e.g.\ $\ket{\psi_1} = c_{\alpha}\ket{\alpha}+c_{\beta}\ket{\beta}$; it has an evolving phase,  $\bra{\alpha}\rho(t)\ket{\beta} = e^{\im (E_{\alpha}-E_{\beta})t/\hbar} \bra{\alpha}\rho(0)\ket{\beta}$, which indicates the chain's coherent dynamics.  For this reason, these off-diagonal terms are called coherences.

\section{The Chain-Bath Coupling\label{chainbathcoupling}}
The first SQUID has a current $I \sigma_1^z$ confined to its ring of cross-sectional area $b$: $\vec{j}_1 =  (I/b) \sigma_1^z \hat{\phi}$ within the ring, $\vec{j}_1=0$ elsewhere.  Here $\hat{\phi}$ is the azimuthal unit vector in cylindrical coordinates centered on the ring.  With
\[
V_1 = (m_e/e) \int d^3r  \, \vec{j}_1\cdot\dot{\vec{u}}
\]
and
\[
\vec{\phi}_{\vec{k}} \equiv \int_0^{2 \pi} d\phi \int_{R-\sqrt{b}/2}^{R+\sqrt{b}/2}dr \, r \int_{-\sqrt{b}/2}^{\sqrt{b}/2} dz\,  \hat{\phi}\, e^{-\im \vec{k}\cdot\vec{r}}
\]
the Fourier transform of the $\hat{\phi}$ within the ring, we have
\[
V_1 =\frac{m_e}{e} \frac{I}{b} \sigma_1^z  \frac{-\im}{\sqrt{V}} \sum_{\vec{k}s}\sqrt{\frac{\hbar \omega_s(\vec{k})}{2 \rho}}\left(a_{\vec{k}s}\vec{\phi}^*_{\vec{k}} - a^{\dagger}_{\vec{k}s} \vec{\phi}_{\vec{k}}\right) \cdot \hat{e}_s(\vec{k})
.\]

In the thin ring approximation, $|\vec{k}| \sqrt{b} \ll 1$, the Fourier transform becomes
\[\vec{\phi}_{\vec{k}} \Rightarrow -\im 2 \pi R b J_1(k_x R) \hat{n}_{\vec{k}},\]
where $J_1(k_x R) = J_1(|\vec{k}| R \sin \theta)$ is the first order Bessel function.  The polar angle $\theta$ is the angle between the $\hat{z}$-axis of the ring and the wavevector $\vec{k}$ while $\hat{n}_{\vec{k}} \perp \vec{k}$ lies in the plane of the ring.   Choosing $\hat{e}_1(\vec{k})$ to lie in the plane of the ring, i.e.\ $\hat{e}_1(\vec{k})=\hat{n}_{\vec{k}}$, we obtain
\[V_1 =    \lambda\, \sigma_1^z  \sum_{\vec{k}} \sqrt{|\vec{k}| R} \, J_1(k_x R) \left(a_{\vec{k}1} \! + a^{\dagger}_{\vec{k}1}\right)
,\]
with coupling constant $\lambda \equiv  2 \pi I (m_e/e) \sqrt{R \hbar c_{\perp}/2 \rho V}$.

The other SQUID rings are centered not at $\vec{r}=0$ but are evenly spaced, a distance $d$ apart, along the $\hat{x}$-axis.  The analysis for each SQUID's coupling to phonons is calculated in its own coordinates $\vec{r}_j$ centered at $\vec{x}_j \equiv d(j-1)\hat{x}$ so that $\vec{r} = \vec{x}_j + \vec{r}_j$.  The creation and annihilation operators' phase factors $e^{\pm\im\vec{k}\cdot\vec{r}}$ become $e^{\pm \im \vec{k} \cdot \vec{x}_j} e^{\pm\im\vec{k}\cdot\vec{r}_j}$ while the rest of the calculations, in the $\vec{r}_j$ coordinates, are exactly the same as before.  The total coupling is thus
\[
V = \lambda \sum_j \sigma_j^z \sum_{\vec{k}}  \sqrt{|\vec{k}| R} \, J_1(k_x R) \left(a_{\vec{k}1}e^{\im \vec{k}\cdot \vec{x}_j}  \!+ a^{\dagger}_{\vec{k}1}e^{-\im \vec{k}\cdot \vec{x}_j}\right)
\!.
\]

\section{The Bath Correlator\label{correlator}}
To evaluate the bath correlator we perform the trace in the bath's energy eigenbasis by summing over the diagonal matrix elements:  $\mbox{Tr}_B [\tilde{\phi}_j(t) \, \tilde{\phi}_k(t') \, \rho_B] =$
\begin{widetext}
\[
\sum_n
\sum_{\vec{k},\vec{k}'}  \sqrt{|\vec{k}|R|\vec{k}'|R} \, J_1(k_x R) \, J_1(k'_x R)
 \bra{n} \left(\tilde{a}_{\vec{k}1}(t)e^{\im \vec{k}\cdot \vec{x}_j}  \!+ \tilde{a}^{\dagger}_{\vec{k}1}(t)e^{-\im \vec{k}\cdot \vec{x}_j}\right)\!\left(\tilde{a}_{\vec{k}'1}(t')e^{\im \vec{k}'\cdot \vec{x}_k}  \!+ \tilde{a}^{\dagger}_{\vec{k}'1}(t')e^{-\im \vec{k}'\cdot \vec{x}_k}\right)\!
\frac{e^{-E_n/k_BT}}{Z_B}\ket{n}
\]
\[
= \overbrace{(V/8\pi^3) \int_0^{\infty} \frac{d\omega\, \omega^2}{c^3_{\perp}}
\int_0^{\pi} d\theta \sin \theta \int_0^{2\pi} d\phi}^{\sum_{\vec{k}}}\, \overbrace{\omega \tau_R \, J_1^2(\omega \tau_R \sin\theta)}^{|\vec{k}|R \,J_1^2(k_x R)
}
 \underbrace{\left[\coth(\frac{\hbar\omega}{2k_BT})\cos(\omega\tau)-\im\sin(\omega\tau)\right]}_{(N_{\omega}+1)e^{-\im\omega\tau}+N_{\omega}e^{\im\omega\tau}}\underbrace{\cos(\omega\tau_{jk}\sin\theta\cos\phi)}_{\Re [e^{\pm\im\vec{k}\cdot(\vec{x}_j-\vec{x}_k}]}
.\]
\end{widetext}
The steps leading to the second line are as follows.  The double sum $\sum_{\vec{k}\vec{k}'}$ collapses to a single sum $\sum_{\vec{k}}$ (which we convert to a $\vec{k}$-space integral in spherical coordinates) because the only non-zero cross-terms $\sim \delta_{\vec{k}\vec{k}'}$.  They are
\[
\bra{n}  \tilde{a}_{\vec{k}1} e^{\im\vec{k}\cdot\vec{x}_j} \tilde{a}^{\dagger}_{\vec{k}'1} e^{-\im \vec{k}'\cdot\vec{x}_k}\ket{n} = (n_{\vec{k}1}+1) \,\delta_{\vec{k}\vec{k}'}  e^{-\im \omega\tau} e^{\im\vec{k}\cdot\vec{x}_{jk}}
\]
and
\[
\bra{n}  \tilde{a}^{\dagger}_{\vec{k}1} e^{-\im\vec{k}\cdot\vec{x}_j} \tilde{a}_{\vec{k}'1} e^{\im \vec{k}'\cdot\vec{x}_k}\ket{n} = n_{\vec{k}1} \, \delta_{\vec{k}\vec{k}'}  e^{\im \omega\tau} e^{-\im\vec{k}\cdot\vec{x}_{jk}},
\]
where the factors $e^{\pm\im\omega\tau}$ arise from being in the interaction picture.  Here we have abbreviated $\omega_1(\vec{k}) = \omega$ for the angular frequency of the mode with wavevector $\vec{k}$ and transverse-in-plane polarization and written $\vec{x}_{jk}=\vec{x}_j-\vec{x}_k$ for the vector connecting the $(j,k)$-pair of SQUIDs.  We have also used $\tau_R \equiv R/c_{\perp}$, which is half the time it takes a phonon to traverse a SQUID.  We switch the order of the sums, $\sum_n\sum_{\vec{k}} \Rightarrow \sum_{\vec{k}}\sum_n$, and write $N_{\omega}\equiv \sum_n n_{\vec{k}1} \frac{e^{-E_n/k_BT}}{Z_B}$ for the thermal-average occupation number.  It sums to $N_{\omega} = 1/(e^{\hbar\omega/k_BT}-1)$:
\[
N_{\omega} =
\underbrace{
\frac
{\sum_{n_{\vec{k}1}} n_{\vec{k}1} e^{-\hbar \omega n_{\vec{k}1}/k_BT}}
{\sum_{n_{\vec{k}1}} e^{-\hbar \omega n_{\vec{k}1}/k_BT}}
}_{1/(e^{\hbar\omega/k_BT}-1)}
\underbrace{
\frac
{\prod_{i\neq\vec{k}1} \sum_{n_i}  e^{-\hbar \omega_i n_i/k_BT}}
{\prod_{i\neq\vec{k}1} \sum_{n_i} e^{-\hbar \omega_i n_i/k_BT}}
}_{1}
.\]
Now every function in the bath correlator besides the $e^{\pm \im \vec{k}\cdot\vec{x}_{jk}}$ is an even function of $\vec{k}$.  The sum over wavevectors thus selects the $\cos (\vec{k}\cdot\vec{x}_{jk})$ part of $e^{\pm \im \vec{k}\cdot\vec{x}_{jk}}$.   Because the $(j,k)$-pair of SQUIDs are positioned on the $\hat{x}$-axis, we use $k_x = |\vec{k}|\sin\theta\cos\phi$ to obtain $\cos(\vec{k}\cdot\vec{x}_{jk})  = \cos(\omega \tau_{jk} \sin\theta\cos\phi)$, with $\tau_{jk}=d(j-k)/c_{\perp}$ the phonon transit time between the SQUIDs.  Finally, it can be shown that $(2 N_{\omega}+ 1) = \coth(\hbar\omega/2 k_BT)$ so that we can write
\[
(N_{\omega}+1)e^{-\im\omega\tau}+N_{\omega}e^{\im\omega\tau} \!=\! \coth(\frac{\hbar\omega}{2k_BT})\cos(\omega\tau)-\im\sin(\omega\tau)
.\]

\section{The Spectral Densities\label{spectraldensities}}

We can write the bath correlator as an integral over bath frequencies, $\mbox{Tr}_B [\tilde{\phi}_j(t) \, \tilde{\phi}_k(t') \, \rho_B] =$
\[
 \underbrace{\int_0^{\infty} \!\!\!\!\! d \omega \, J_{jk}(\omega) \coth(\frac{\hbar\omega}{2 k_BT}) \cos(\omega \tau)}_{\nu_{jk}(\tau)} \!- \im \!\! \underbrace{\int_0^{\infty} \!\!\!\!\!  d \omega \, J_{jk}(\omega)\sin(\omega \tau)}_{\mu_{jk}(\tau)}
,\]
characterized by spectral densities
\[
J_{jk}(\omega) = \frac{V}{8\pi^3} \, \frac{\tau_R}{c_{\perp}^3} \, \omega^3 \, \Theta_{jk}(\omega\tau_R)
\]
for which we need the angular integration
\[
\Theta_{jk}(\omega \tau_R) \equiv \! \int_0^{\pi} \!\! d\theta\, \sin \theta J_1^2(\omega\tau_R\sin\theta) \, \!\!\!\!\!\! \underbrace{2\pi J_0(\omega \tau_{jk} \sin\theta)}_{\int \!\! d\phi\, \cos(\omega\tau_{jk}\sin\theta\cos\phi)} \!\!\!\!\!.
\]

For the single-SQUID ($j=k$) $\theta$ integration we obtain
\[
J_{jj}(\omega) = \frac{\hbar m_e^2 I^2 R^4 \omega^5}{6 \lambda^2 e^2 \rho c_{\perp}^5} {}_pF_q(\{\frac{3}{2}\},\!\{\frac{5}{2},3\},\!-(\omega \tau_R)^2)\]
independent of $j$ and with ${}_pF_q$ the generalized hypergeometric function.   Although we use these exact ``${}_pF_q$'' spectral densities in our numerical simulations, it is helpful to know that for a ``small ring'' ($\omega \tau_R \ll 1$) the single small SQUID spectral density is approximately
\[
J^{(S)}_{jj}(\omega) \Rightarrow \frac{2 \hbar m_e^2}{3 \lambda^2 \pi^2 e^2}\frac{I^2(\pi R^2)^2}{4 \rho c_{\perp}^5}\omega^5\]
while for a ``large ring'' ($\omega \tau_R \gg 1$) the single large SQUID spectral density is approximately
\[
J^{(L)}_{jj}(\omega) \Rightarrow \frac{2 \hbar  m_e^2}{\lambda^2 e^2}\frac{I^2 R}{4\rho c_{\perp}^2}\omega^2
.\]

Next, for $\omega\tau_{jk}\ll1$, as can happen for ``large'' SQUIDs at a level crossing where $\omega\approx 0$ or, regardless of $\omega$, for a few small SQUIDs spaced only a few small SQUID radii apart, the cross term spectral densities must be considered.  It turns out that to first order in $\omega\tau_{jk}$ we can use the single SQUID $j=k$ rates for the $j\neq k$ cross terms, as follows.  $\omega\tau_{jk}\ll1$ implies $\omega\tau_R \ll1$ so we reconsider our $\theta$ integrations with $J_1(\omega\tau_R\sin\theta) \approx \frac{1}{2} \omega\tau_R\sin\theta$ to obtain
\[
\frac{\Theta_{jk}(\omega\tau_R)}{\omega^2\tau_R^2/4} \Rightarrow \overbrace{4\pi \frac{\omega\tau_{jk}\cos(\omega\tau_{jk})\!-\!(1\!-\!\omega^2\tau_{jk}^2)\sin(\omega\tau_{jk})}{\omega^3\tau_{jk}^3}}^{2 \pi \int d\theta \sin^3 \theta J_0(\omega\tau_{jk}\sin\theta)}
,\]
which, to first order in $\omega\tau_{jk}$, give spectral densities identical to those of a single small SQUID
\[
J^{(S)}_{j\neq k} \Rightarrow J^{(S)}_{jj}(\omega) {\mbox{ when }} \omega \tau_{jk} \ll1
\]
and we may as well use the equivalent  ${}_pF_q$ spectral densities.

Finally, we assume that when $\omega\tau_{jk}\gg1$ we can neglect the cross term ($j\neq k$) spectral densities.  The assumption relies on the $J_0(\omega\tau_{jk}\sin\theta)$ kernel of the $\theta$ integration oscillating quickly enough between positive and negative values that the integral never accumulates any significant value.

Again, the ${}_pF_q$ spectral densities are the ones we use for $j=k$ {\em and} for close-together or far-apart SQUIDs satisfying $\omega\tau_{jk}\ll1$.   For far-apart SQUIDs satisfying $\omega\tau_{jk}\gg 1$ we use $J_{j\neq k} (\omega) = 0$.

\section{Constant (Markov) Coefficients\label{constantcoefficients}}
The coefficients of diffusion, renormalization, anomalous diffusion, and damping are obtained by convolving a $\cos\omega\tau$ or $\sin\omega\tau$ with the $\nu_{jk}(\tau)$ or $\mu_{jk}(\tau)$ kernels.  But the only time-dependent terms in the integrands for the kernels are $\cos\omega'\tau$ or $\sin\omega'\tau$.  Since these time-convolutions are relatively simple, in our calculation of the coefficients we switch the order of the integrals, $\int d\tau \int d\omega' \Rightarrow \int d\omega' \int d\tau$.

For the diffusion and damping we find that the integrations $\int_0^t d\tau \cos\omega'\tau\cos\omega\tau$ or $\int_0^t d\tau \sin\omega'\tau\sin\omega\tau$ are
\[
\frac{1}{2}\left[\frac{\sin(\omega'-\omega)t}{\omega'-\omega} \pm \frac{\sin(\omega'+\omega)t}{\omega'+\omega} \right]
,\]
which behave, as $t\rightarrow\infty$, like Dirac delta functions:  $\frac{\pi}{2}\left[\delta(\omega'-\omega) \pm \delta(\omega'+\omega)\right]$.  This is because they oscillate with $\omega'$ at a frequency $t$ everywhere except at $\pm \omega$, where they spike to a height $\sim t$ and width $\sim 1/t$.  As long as the frequency of these oscillations is much faster than any features of the spectral densities (i.e.\ $t\gg\tau_{jk},\tau_R$), of the cutoff (i.e.\ $t\gg 1/\Lambda$), and of the hyperbolic cotangent (i.e.\ $t \gg \hbar/k_BT$) then the only contribution to the $\omega'$ integral comes from the spikes at $\pm \omega$.  Subject to these rough criteria,
\[
\begin{array}{ccccc}
t\gg\tau_{jk},\tau_R & &  t \gg 1/\Lambda & & t \gg \hbar/k_BT\\
\end{array}
,\]
the coefficients of diffusion and damping approach constant values which, thanks to the Dirac delta functions, are easy to identify:
\[
D^{\omega}_{jk}  \equiv \frac{\pi}{2} J_{jk}(\omega) \coth(\frac{\hbar\omega}{2 k_BT})  = \lim_{t\rightarrow\infty} D^{\omega}_{jk}(t)
\]
and
\[
\gamma^{\omega}_{jk}  \equiv \frac{\pi}{2} J_{jk}(\omega)  = \lim_{t\rightarrow\infty} \gamma^{\omega}_{jk}(t)
.\]

For the renormalization and anomalous diffusion, the integrations $\int_0^t d\tau \sin\omega'\tau\cos\omega\tau$ or $\int_0^t d\tau \cos\omega'\tau\sin\omega\tau$ are
\[
\frac{\sin^2\frac{\omega'+\omega}{2}t}{\omega'+\omega} \pm \frac{\sin^2\frac{\omega'-\omega}{2}t}{\omega'-\omega}
\]
whose behavior for $t\rightarrow\infty$ is not so clear.  However, for large $t$ these terms oscillate so fast with $\omega'$ that they too average away the time dependence and we can at least define, using an explicit exponential cutoff $e^{-\omega'/\Lambda}$, constant values for the renormalization,
\[
r_{jk}^{\omega} \equiv \lim_{t\rightarrow\infty} \int_0^{\infty} \!\!\!\!\!  d\omega' J_{jk}(\omega') e^{-\omega'/\Lambda}
\left[\frac{\sin^2\frac{\omega'+\omega}{2}t}{\omega'+\omega} + \frac{\sin^2\frac{\omega'-\omega}{2}t}{\omega'-\omega}\right]
,\]
and anomalous diffusion,
\[
A_{jk}^{\omega} \equiv \int_0^{\infty} \!\!\!\!\!  d\omega' J_{jk}(\omega') \coth(\frac{\hbar\omega'}{2k_BT})e^{-\omega'/\Lambda}
\!\! \int_0^{t\rightarrow\infty} \!\!\!\! \!\!\!\! \!\!\! d\tau  \cos\omega'\tau\sin\omega\tau
,\] which we use symbolically without actually ever evaluating them.

\bibliography{squids}

\end{document}